\documentclass{elsart}
\usepackage{epsf}
\usepackage{amsmath}
\usepackage{amssymb}
\usepackage{bbm}

\newcommand{\erg}[1]{E_{\mathrm{#1}}}
\newcommand{\sig}[1]{\sigma_{\mathrm{#1}}}
\newcommand{\cov}[1]{\eta_{\mathrm{#1}}}

\newcommand{\Erg}[1]{E^{\mathrm{#1}}}
\newcommand{\num}[1]{n^{\mathrm{#1}}}
\newcommand{\pot}[1]{\mu^{\mathrm{#1}}}

\begin{document}

\begin{frontmatter}

\title{Interplay of Strain Relaxation and Chemically Induced Diffusion Barriers:
Nanostructure Formation in 2D Alloys}

\author[wuerzburg]{T. Volkmann\corauthref{cor}},
\author[wuerzburg]{F. Much},
\author[groningen]{M. Biehl},
\author[prague]{M. Kotrla}

\address[wuerzburg]{
Institut f\"{u}r Theoretische Physik und Astrophysik,
Universit\"{a}t W\"{u}rzburg, Am Hubland, 97074 W\"{u}rzburg, Germany
}

\address[groningen]{
Institute of Mathematics and Computing Science,
University of Groningen, P.O. Box 800, 9700 AV Groningen, The Netherlands
}

\address[prague]{
Institute of Physics,
Academy of Sciences of the Czech Republic, Na Slovance 2, 182 21 Prague 8, Czech Republic
}

\corauth[cor]{Corresponding author. Tel.: +49-931-888-5733; 
Fax: +49-931-888-5141; Email: volkmann@physik.uni-wuerzburg.de}


\begin{abstract}

We study the formation of nanostructures with alternating stripes
composed of bulk-immiscible adsorbates during submonolayer heteroepitaxy.
We evaluate the influence of two mechanisms considered in the literature:
(i) strain relaxation by alternating arrangement of the adsorbate species, 
and (ii) kinetic segregation due to chemically induced diffusion barriers. 
A model ternary system of two adsorbates with opposite misfit relative to 
the substrate, and symmetric binding is investigated by off-lattice as well 
as lattice kinetic Monte Carlo simulations.
We find that neither of the mechanisms (i) or (ii) alone 
can account for known experimental observations.
Rather, a combination of both is needed.
We present an off-lattice model 
which allows for a qualitative reproduction
of stripe patterns as well as island ramification 
in agreement with recent experimental observations for
CoAg/Ru(0001) [R.\ Q.\ Hwang, Phys.\ Rev.\ Lett.\ 76, 4757 (1996)]. 
The quantitative dependencies of stripe width and degree of island
ramification on the misfit and interaction strength between the two 
adsorbate types are presented.
Attempts to capture essential features in a simplified lattice gas model show
that a detailed incorporation of non-local effects is required.

\end{abstract}


\end{frontmatter}


\section{Introduction}
\label{introduction}

 Heteroepitaxial growth of thin films 
 has been a field of growing interest in recent years \cite{herman:epitaxy:2004}
 as it displays a variety of highly non-trivial phenomena. 
 Among these are, e.g., the self-organized formation of three-dimensional islands, 
 so-called Quantum Dots \cite{joyce:quantumdots}, 
 self-assembly of ordered nanoscale domain patterns \cite{plass:jpcm14}
 or lateral multilayers \cite{tober:prl81}, or the emergence of 
 misfit dislocations \cite{thayer:prl86}.
 New kinds of materials with unique properties have been fabricated and
 numerous technical applications are based on hetero-systems. 
 This includes, to name only a few, laser diodes, solar cells, and magnetic 
 or magneto-optical storage devices.

 Besides the technological relevance, heteroepitaxy is
 highly interesting from the theoretical point of view.
 It provides a workshop to develop and put forward
 novel approaches and simulation techniques which go
 beyond the more frequent modeling of homoepitaxial systems \cite{kotrla:atomistic:2002}.
 In particular, the correct treatment of kinetic effects in strained
 systems calls for the development of multiscale techniques.
 Despite increasing activity in this direction our present understanding of 
 heteroepitaxy on the microscopic level remains rather limited.

 In the context of metal epitaxy, the formation of surface alloys
 is of particular interest. In many cases, adsorbate and substrate
 intermix and form a thin film of alloy \cite{tersoff:prl74}.
 Another interesting observation is that the deposition
 of two bulk-immiscible metals, say ``A'' and ``B'', upon a suitable substrate ``S'' 
 can result in the formation of a two-dimensional 
 A--B alloy in the first or several layers of adsorbate.
 The ordered nanoscale structure formed by alternating domains 
 of material A and B is of particular interest. 
 Such domains are vaguely called stripes or veins, and their formation
 with a width on the order of nanometers has been observed in a variety of 
 AB/S material systems, including
 CoAg/Ru(0001) \cite{hwang:prl76,hwang:chemrev97},
 CoAg/Mo(110) and FeAg/Mo(110) \cite{tober:prl81}, CuAg/Ru(0001)
 \cite{stevens:prl74}, or PdAu/Ru(0001) \cite{sadigh:prl83}.
 Besides the above mentioned stripe substructure,
 the two component islands in some cases also display dendritic growth 
 \cite{hwang:prl76,hwang:chemrev97}.

 In this paper, we investigate microscopic mechanisms 
 relevant for the self-organized formation of nanoscale features
 in a model ternary AB/S material system during submonolayer growth.
 Our model system accounts for the key characteristics of the above examples,
 namely that the atomic size of adsorbate material A is smaller than that of 
 the substrate S whereas that of adsorbate B is larger.
 One expects that the presence of both
 positive and negative misfit in the same heteroepitaxial system
 will play an important role in the formation and the detailed structure
 of the growing film.
 Furthermore, we have to take into account differences in
 binding energies, and also that the structures are prepared by growth,
 i.e.\ under non-equilibrium conditions.
 On the other hand, interdiffusion of substrate and adsorbates can essentially
 be neglected in these systems.

\begin{figure}[t]
   \begin{minipage}{0.99 \textwidth}
     \epsfxsize= 0.49\textwidth
     \epsffile{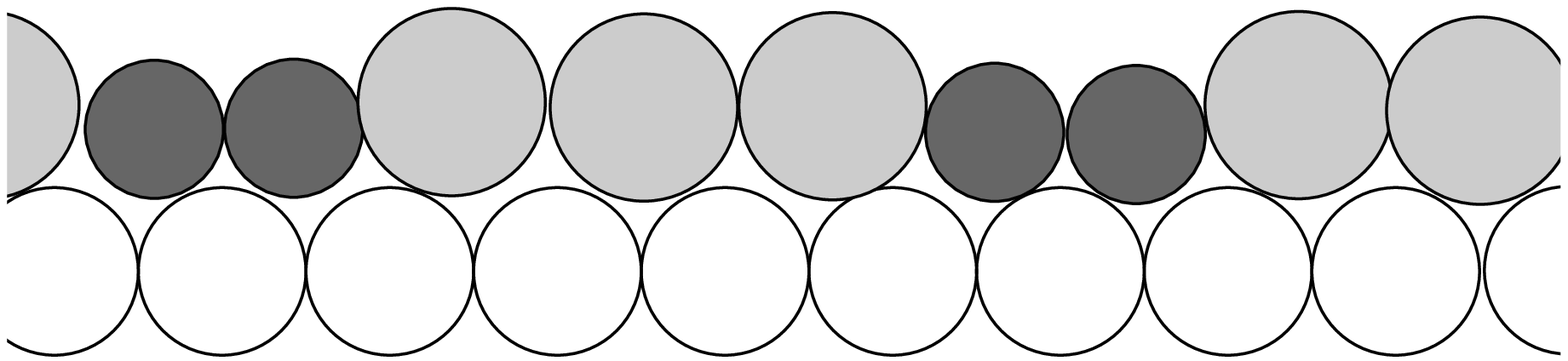}
   \end{minipage}
   \caption{\label{misfitbild}
     Illustration of a ternary system of bigger 
     B-particles (light gray) and smaller A-particles 
     (dark gray) on a substrate (white) of
     intermediate lattice spacing. The effective misfit of the
     adsorbate film can be reduced by an alternating arrangement of
     the species.
   }
 \end{figure}

\begin{figure}[t]
   \begin{minipage}{0.99 \textwidth}
     \epsfxsize= 0.49\textwidth
     \epsffile{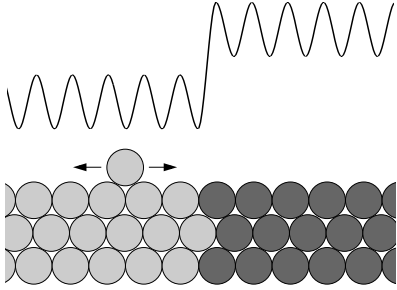}
   \end{minipage}
   \caption{\label{barrierenbild}
     Illustration of a chemically induced step edge barrier.
     Lower part: top view on the step
     edge of an island composed of A-particles (dark gray) and B-particles (light gray). 
     Upper part: schematic diagram of the energy experienced
     by a B-particle diffusing along the step edge.
   }
 \end{figure}

 Mainly two mechanisms were discussed in the literature in 
 the context of stripe formation:

 \begin{itemize}
 \item[a)] Strain relaxation, see e.g.\ \cite{tober:prl81}. \\
   As the misfit of A/B particles is negative/positive with respect
   to the substrate, it is possible to achieve a low effective
   adsorbate misfit by an alternating arrangement of the species.
   The essentially geometric effect is illustrated in
   Fig.\ \ref{misfitbild}.
 \item[b)] Kinetic segregation, see e.g.\ \cite{hwang:prl76}.  \\
   If we assume that the inter-species binding A--B 
   is weaker than that of A--A and B--B, the system tends to separate
   the elements with a boundary as short as possible.
   Clearly, in non-equilibrium island growth this cannot be achieved.
   However, the different binding energies can result in a strong
   kinetic effect for diffusion along existing edges: 
   a B particle, say, is subject to an extra barrier for
   diffusion hops from a B to an A domain and vice versa, cf.\ Fig.\ \ref{barrierenbild}.
   Hence, A and B adatoms will preferentially
   contribute to the growth of domains containing the same
   species.
 \end{itemize}

 Both effects might be sufficient to explain certain aspects of
 the observed non-equilibrium structures.
 The main aim of this work
 is to clarify their role and potential competition in the
 process of island and stripe formation.  
 With the help of atomistic simulations,
 we will demonstrate that both mechanisms are 
 indeed relevant, and that it is their
 interplay which determines the precise film structures.


The paper is organized as follows. In Sec.\ \ref{continuous-description}, we provide
a continuous description of a model ternary system, using an off-lattice simulation model 
(Sec.\ \ref{off-lattice-model}) which incorporates both misfit-induced strain and binding 
energy effects. The behavior of the model is studied under both equilibrium 
(Sec.\ \ref{equilibrium}) and non-equilibrium growth conditions 
(Sec.\ \ref{non-equilibrium}).
The influence of misfit and binding energies on the resulting morphologies is 
discussed. 
In Sec.\ \ref{lattice-description}, the off-lattice simulations
are followed by a description within the framework of the lattice gas method.
In order to determine the role of kinetic effects separately from strain effects,
a lattice  model which incorporates the basic 
difference in the binding energies of adsorbate species but 
lacks an explicit representation of strain is introduced (Sec.\ \ref{lattice-gas-model}).
In Sec.\ \ref{simplified-model} a simplified version of the model which treats 
both adsorbate species in a symmetric way is investigated and the influence of the binding
energies is discussed. 
In order to compare off-lattice and lattice descriptions, a modified 
version of the lattice gas model with parameters fitted to characteristic 
off-lattice diffusion barriers is investigated in Sec.\ \ref{modified-model}.
Section \ref{summary} summarizes and discusses the obtained results and 
Sec.\ \ref{conclusion} gives a conclusion.


\section{Continuous description}
\label{continuous-description}

\subsection{Off-lattice simulation model}
\label{off-lattice-model}

In order to simulate heteroepitaxial growth of an adsorbate on a chemically different substrate 
it is necessary to overcome the limitations of a pre-defined lattice as is discussed, e.g., in 
\cite{schindler:diss:1999,much:diss:2003}.
For this reason we use a recently introduced off-lattice model
\cite{much:epl56} which was shown to successfully
describe a variety of phenomena observed in heteroepitaxial growth, 
including dislocation formation, 
wetting layer and island formation in the Stranski-Krastanov growth mode 
\cite{much:epl56,much:cpc147,much:epl63}.
For a detailed overview, see \cite{much:diss:2003}.
In this model two particles 
which are separated by a {\em continuous\/} distance $r$ interact via
a simple pair-potential $U(r)$, an example being the Lennard-Jones (LJ) potential
\begin{equation}\label{LJ_612}
  U_{\mathrm{LJ}}(r) =\ 4\,E \left[\left(\frac{\sigma}{r}\right)^{\! 12}
    -\left(\frac{\sigma}{r}\right)^{\! 6}\right],
\end{equation}
where $E$ determines the depth of the potential and the equilibrium distance between 
two isolated particles is given by $r_0 = \sqrt[6]{2}\,\sigma$. 
By appropriate choice of the parameters $E$ and $\sigma$, different material properties may be 
specified in the model qualitatively.
For example, interactions between two substrate or adsorbate particles are governed by the sets 
$\{\erg{S},\sig{S}\}$ and $\{\erg{A},\sig{A}\}$, respectively. 
To keep the number of parameters small the standard choice
$\erg{AS}=\sqrt{\erg{A} \erg{S}}$, $\sig{AS}=(\sig{A}+\sigma_{\mathrm{S}})/2$ is used for the 
interaction between adsorbate and substrate particles. 
Since the lattice spacing in a Lennard-Jones crystal is proportional to $\sigma$ 
\cite{ashcroft:ssp:1976}
the relative lattice misfit $\varepsilon$ in the model may directly be controlled by the values of 
$\sigma_{\mathrm{S}}$ and $\sig{A}$:
\begin{equation}\label{misfit}
  \varepsilon = \frac{\sig{A}-\sigma_{\mathrm{S}}}{\sigma_{\mathrm{S}}}.
\end{equation}
In our previous work we have addressed rather fundamental aspects of heteroepitaxial growth
\cite{much:epl56,much:cpc147,much:epl63} instead of focusing on specific material properties.
In order to save computer time, the simulations therefore were done in $1+1$ dimensions.
However, phenomena like the formation of alternating vein structures cannot
be mapped to $1+1$ dimensions. 
For this reason, we will extend the simulation method to $2+1$ dimensions, here.
In order to keep the computational effort acceptable we choose a simple cubic (sc) 
lattice symmetry for our simulations. 
The advantage is that due to the lower coordination number less particles have to be taken 
into account for energy calculations than in a close-packed lattice.
Note that the majority of the experimental results discussed in Sec.\ \ref{introduction}
are for metals grown on substrates with fcc/hcp symmetry. 
However, this difference should primarily affect the geometry of surface features.
We believe that our qualitative conclusions will not depend on this 
simplification.

In order to stabilize the sc lattice, we adapt the method proposed in \cite{schroeder:surf375} 
and choose
\begin{equation}
  \label{CUBIC}
  V(r) = \left(0.1 + 8\,\left( \frac{x^2}{r^2}-\frac{1}{2} \right)
    \left( \frac{y^2}{r^2}-\frac{1}{2} \right)
    \left( \frac{z^2}{r^2}-\frac{1}{2} \right)\right)U(r)
\end{equation}
as interaction potential between two particles separated by a distance $r$.
Two kinds of pair-potentials $U(r)$ are used:
the LJ potential given by Eq.\ (\ref{LJ_612}) and the Morse potential
\begin{equation}\label{MORSE}
  U_{\mathrm{M}}(r) = E\,e^{a\left({\sigma}-r\right)} 
  \left(e^{a\left({\sigma}-r\right)}-2 \right).
\end{equation}
Similar to the LJ potential, the depth of the Morse potential is given by $E$,  
and the equilibrium distance between two isolated particles becomes $r_0 = \sigma$.
The additional parameter $a$ in Eq.\ (\ref{MORSE})
determines the steepness of the Morse potential around its minimum.
In our simulations we use $a=5.0$, $5.5$ and $6.0$, corresponding to an increase of the steepness.
In order to save computer time, $U(r)$ is cut off for 
particle distances greater than $r_{cut} = 2\,r_{0}$ during energy calculations, 
whereas for the calculation of 
diffusion barriers the cut-off distance is set to $3\,r_{0}$. 
These simplifications are perfectly justified since both the LJ and the Morse 
potential decline fast towards zero with increasing particle distance.

In the following we consider two different adsorbate types, called A and B, 
with negative and positive misfit, respectively, relative to a substrate S. 
The interaction strength between two substrate particles is given by
$\erg{S}$ and $\sigma_{\mathrm{S}}=1$ whereas $\erg{A}$, $\sig{A}$ and $\erg{B}$, $\sig{B}$ 
are chosen for A--A and B--B interactions, respectively.
For the interaction between adsorbate particles of type X $\in$ \{A,B\} and the substrate we use
$\erg{XS}=\sqrt{\erg{X} \erg{S}}$ and $\sig{XS} = (\sig{X}+\sig{S})/2$ whereas
$\erg{AB}$ and $\sig{AB} = (\sig{A}+\sig{B})/2$ hold for the interaction between 
A and B adsorbate particles.
The misfit is assumed to be symmetric in the system:
\begin{equation}\label{symmetric}
 \sig{A} = 1-\varepsilon \quad \mathrm{and} \quad \sig{B} = 1+\varepsilon
\end{equation}
with $\varepsilon > 0$. 
Although experimental systems fulfill this symmetry only approximately we do not 
expect this to be crucial and 
restrict ourselves to a single parameter $\varepsilon$. 
\begin{table}
  \caption{\label{TAB_ES}The substrate-substrate interaction $\erg{S}$ used in the Lennard-Jones 
    potential (LJ) and the Morse potential with parameter $a$ (M$_a$).
  }
  \begin{tabular}{lcccc} \hline\hline
    & LJ & M$_{5.0}$ & M$_{5.5}$ & M$_{6.0}$  \\ \hline
    $\erg{S}\,[\mathrm{eV}]$ & 3.0 & 3.0  & 2.814 & 2.70 \\ \hline\hline
  \end{tabular}
\end{table}
The potential depths are chosen in such a way that they meet two demands:
on the one hand the ratio between $\erg{S}$ and $\erg{A}$, $\erg{B}$ is kept fixed for all 
potentials,
\begin{equation}\label{E_S}
  \erg{A} = \erg{B} = \frac{1}{6} \erg{S},
\end{equation}
and is chosen such that substrate particles are bound 
much more strongly and thus intermixing of adsorbate and substrate particles is suppressed. 
On the other hand, in the case of homoepitaxy ($\varepsilon=0$) the diffusion barrier 
on plain substrate $E_{a,sub}$ should have roughly the same value for all used potentials 
to facilitate the comparison of the results.
We choose $\erg{S}$ here in such a way that for homoepitaxy 
$E_{a,sub} \approx 0.37\,\mathrm{eV}$ --- a typical value for self-diffusion barriers of metals
(see e.g.\ \cite{trushin:surf389,maca:vacuum54,yu:prb56,kellog:prl64}).
The resulting $\erg{S}$ for the different potentials are listed in Table \ref{TAB_ES}.


\subsection{Equilibrium simulations}
\label{equilibrium}

\begin{figure}[t]
  \begin{minipage}{0.99 \textwidth}
    \epsfxsize= 0.99\textwidth
    \epsffile{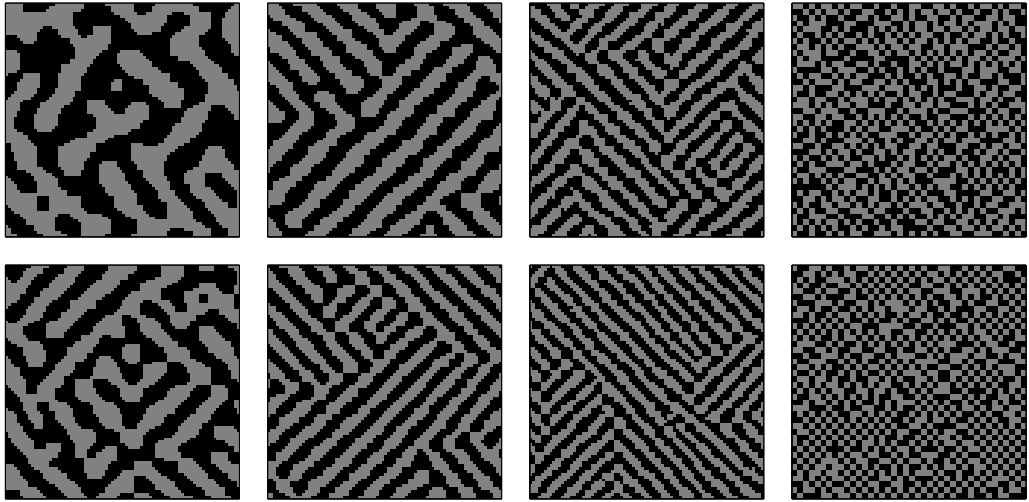}
  \end{minipage}
  \caption{\label{FIG_LJ_SNAPSHOTS}
    Snapshots for equilibrium simulations with the Lennard-Jones potential at $T=250\,\mathrm{K}$ 
    for
    $\erg{AB} = 0.6 \erg{A}, 0.8 \erg{A}, 0.9 \erg{A}, 1.0 \erg{A}$ (from left to right) and 
    $\varepsilon = 4.5\,\%$ (top), 
    $\varepsilon = 5.5\,\%$ (bottom). The particle concentrations
    are $\cov{A} = \cov{B} = 0.5$. The panels for
    $\erg{AB} = 1.0 \erg{A}$ show $40 \times 40$ sections, 
    the remaining panels $80 \times 80$ sections
    of the $100 \times 100$ system. The bigger B particles appear in light gray.}
\end{figure}

In order to determine the influence of misfit 
and binding energy between A and B particles 
on the resulting surface patterns,
we carry out canonical equilibrium simulations with a fully covered substrate and fixed 
concentrations $\cov{A}$, $\cov{B}$ of A and B particles ($\cov{A}+\cov{B}=1$).
The substrate is prepared as a six-layer-thick crystal with $100 \times 100$ particles 
in each layer and fixed particle positions in the bottom layer.
Periodic boundary conditions are applied in the $x$- and $y$-direction. 
For the range of misfits $\varepsilon$ considered in our simulations we do not observe
the formation of dislocations even at full coverage.
The continuous $x$- and $y$-positions of any given adsorbate particle are thus close to 
the coordinates of a distinct lattice site in a $100 \times 100$ square lattice with
discrete sites. 
At the beginning of each simulation run
the substrate is randomly covered with adsorbate particles with a given ratio $\cov{A}/\cov{B}$. 
Then the system is driven towards thermal equilibrium at temperature $T$ by means of 
a rejection-free algorithm \cite{newman:montecarlo:1999}
where A and B particles are exchanged \cite{much:diss:2003,ahr:surf505}.
Since, here, we are not interested {\em how\/} the system approaches equilibrium we 
choose a nonlocal dynamics where
the range of particle jumps is unlimited. This yields considerably
faster equilibration compared to local Kawasaki-type dynamics \cite{newman:montecarlo:1999}.
In each event an A particle at site $i$ of the square lattice exchanges its binding site with 
a B particle at site $j$ according to the rate
\begin{equation}\label{rate_equilibrium}
  r_{i \rightarrow j} = \exp\left(\frac{\Delta H_i - \Delta H_j}{2 k_B T} \right)
\end{equation}
where $\Delta H_x = H_x(\mathrm{A}) - H_x(\mathrm{B})$ gives the energy difference of the system 
with site $x$ occupied with an A or B particle. 
$H_x(\mathrm{A})$ and $H_x(\mathrm{B})$ are calculated in a local way: 
an A particle is set to site $x$ and all particles
within $r_{cut} = 2\,r_0$ around this site are allowed to relax locally. 
The local energy is registered as $H_x(\mathrm{A})$. 
In a similar way we obtain $H_x(\mathrm{B})$. 
Thus, the rates given by Eq.\ (\ref{rate_equilibrium}) 
fulfill the detailed balance condition.
To avoid complications in the calculation of
the configurational energies $H_x(\mathrm{A})$ and $H_x(\mathrm{B})$
we permit only exchanges 
between sites $i$ and $j$ which are more than $r_{cut}$ away from each other.

In order to avoid accumulation of artificial strain due to the local relaxation for the
calculation of $\Delta H_x$, the system is globally relaxed after a fixed number of 
simulation steps (here $5000$) and all rates are re-evaluated.
The system's total energy is registered after each
global relaxation. All simulation runs are halted after $20$ global relaxation events, 
i.e.\ after $10^5$ elementary simulation steps.

Figure \ref{FIG_LJ_SNAPSHOTS} shows simulation results for the cubic
LJ potential [Eqs.\ (\ref{LJ_612}), (\ref{CUBIC})] for two different values of 
the misfit $\varepsilon$
and various strengths of the A--B interaction $\erg{AB}$. The particle concentrations are 
$\cov{A} = \cov{B} = 0.5$.
For each parameter set a regular arrangement of alternating
A and B stripes may be identified, which are oriented along the $\left<11\right>$ directions, 
preferentially. As known
from other atomistic models with size mismatch \cite{tersoff:prl74,krack:prl88}
these regular patterns arise from the competition between binding energy of the particles and 
strain energy.
As one can see in Fig.\ \ref{FIG_LJ_SNAPSHOTS}, with increasing $\erg{AB}$ and increasing 
$\varepsilon$ the stripes
become thinner and more regular in size and shape. For the case $\erg{AB} = \erg{A} = \erg{B}$ 
the system approaches a checkered state, i.e.\ a stripe width of one. 
The alignment of the stripes along the $\left<11\right>$ directions is due to
the cubic symmetry of the potential: both particle types try to reach their preferred 
stripe width in each lattice direction ($x$ and $y$). 
Note, that the used cubic form of the potential [Eq.\ (\ref{CUBIC})] has only a
weak interaction in the $\left<11\right>$ direction \cite{much:diss:2003}.

\begin{figure}[t]
  \begin{minipage}{0.47 \textwidth}
    \epsfxsize= 0.99\textwidth
    \epsffile{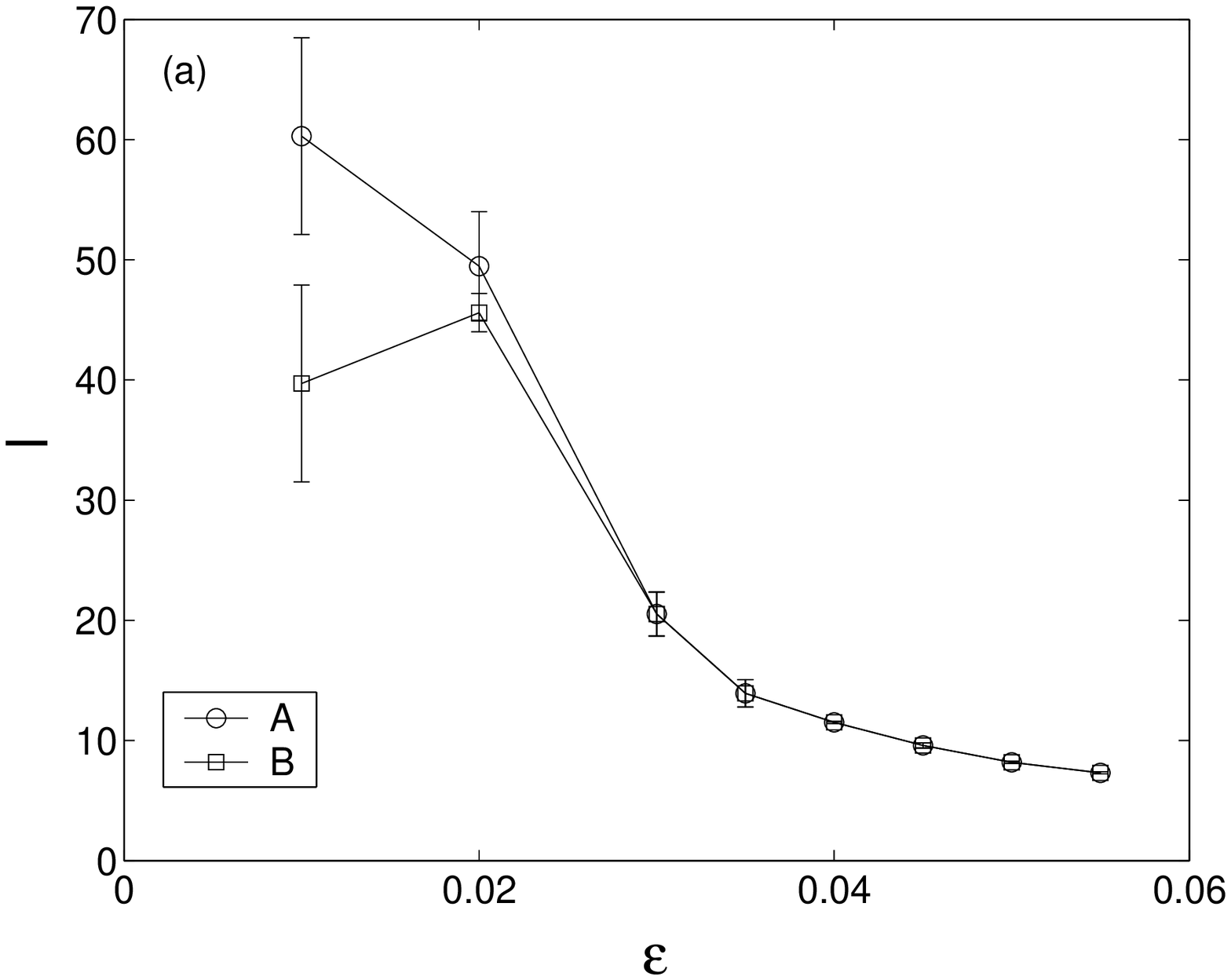}
  \end{minipage}
  \hfill
  \begin{minipage}{0.45 \textwidth}
    \epsfxsize= 0.99\textwidth
    \epsffile{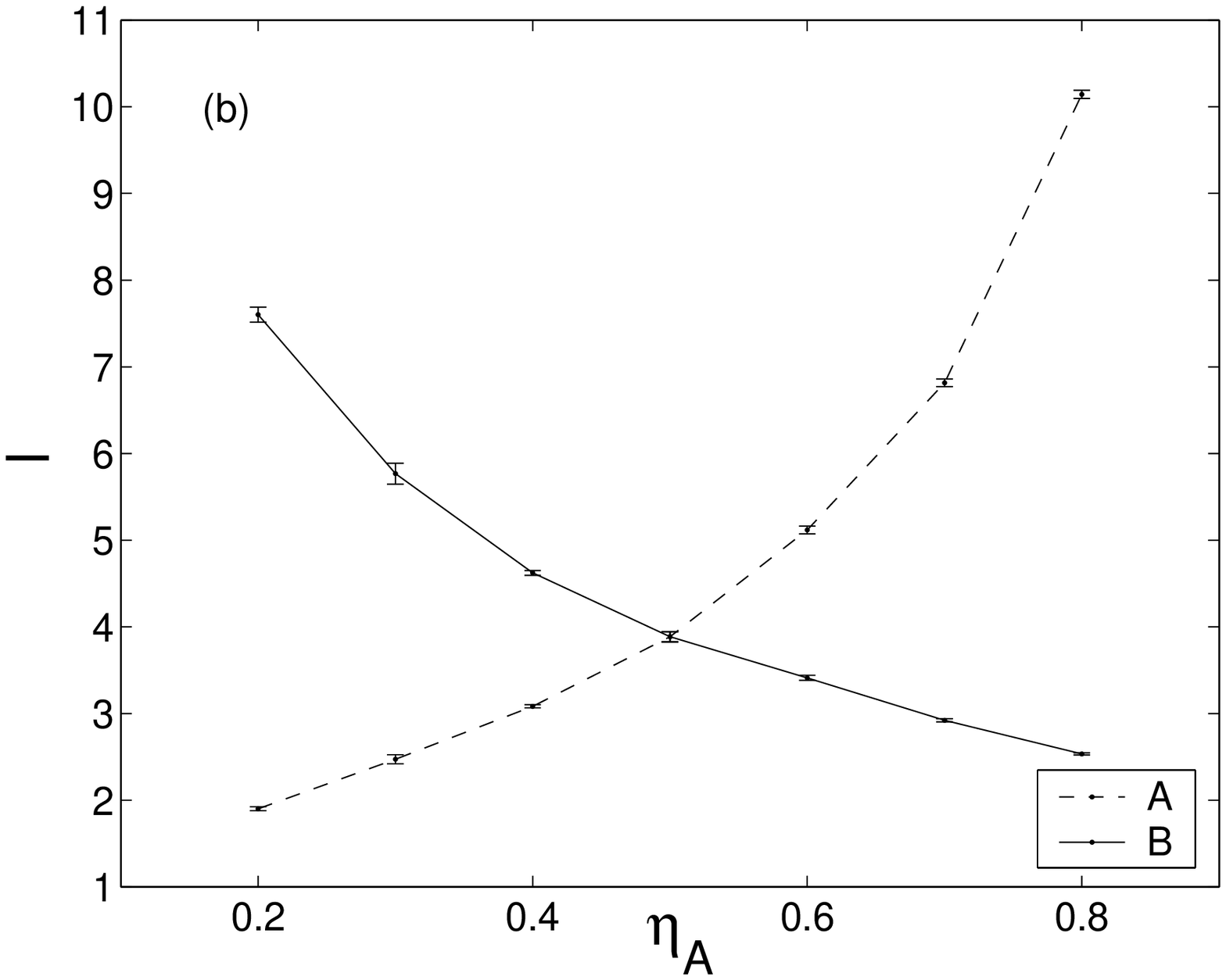}
  \end{minipage}
  \caption{\label{LEN_W1}Equilibrium simulations with the Lennard-Jones potential at 
    $T=250\,\mathrm{K}$.
    (a) Misfit dependence of the stripe widths for $\erg{AB}=0.6\erg{A}$ and particle 
    concentrations $\cov{A}=\cov{B}=0.5$.
    Due to the onset of stripe formation along $\left<10\right>$
    the determination of the stripe width becomes inaccurate for misfits $\varepsilon\leq 0.01$.
    (b) Stripe widths for $\erg{AB}=0.9\erg{A}$ and $\varepsilon=5\%$ in dependence of 
    the A particle
    concentration $\cov{A}$ ($\cov{B}=1-\cov{A}$, consequently).
    Each value is obtained by averaging over three independent simulation runs.}
\end{figure}

Figure \ref{LEN_W1}(a) shows the width $l$ of A and B stripes for $\erg{AB} = 0.6\,\erg{A}$ 
in dependence of
the misfit. Since the concentrations of A and B particles are equal
the stripes have about the same width for both adsorbate types.
For very small misfits the alignment of the stripes along $\left< 11 \right>$ vanishes in favor 
of a $\left< 10 \right>$
orientation which decreases the interfacial energy between A and B regions.
This process is reflected in the large deviations of the stripe width at $\varepsilon = 0.01$ in
Fig.\ \ref{LEN_W1}(a).

The situation changes completely for $\cov{A} \not= \cov{B}$.
As Fig.\ \ref{LEN_W1}(b) shows for $\erg{AB}=0.9\erg{A}$ and $\varepsilon=5\%$, the stripe width 
increases with
increasing concentration of the particle type. It is noticeable that the bigger B particles form
thinner stripes at high B concentration than the smaller A particles at high A concentration.
This is due to the asymmetric pair-potential, which is steeper in compression than in tension 
and thus
(compressed) B stripes are slightly more restricted in their width than A stripes.

With otherwise unchanged parameters we performed additional simulations for 
the Morse potential with $a = 6.0$, which is
steeper in both---compression and tension---than the LJ potential used before.
However, LJ and Morse potential yield
quite similar results: again the competition between strain and binding energy causes 
alternating stripes of
decreasing width with increasing $\varepsilon$. Due to the cubic symmetry the stripes are again
solely aligned in the $\left< 11 \right>$ direction, only for very small misfits stripes 
can also be found along $\left< 10 \right>$.

\begin{figure}[t]
  \begin{minipage}{0.45 \textwidth}
    \epsfxsize= 0.99\textwidth
    \epsffile{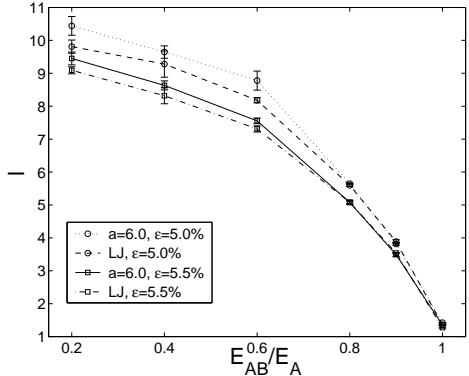}
  \end{minipage}
  \caption{\label{LEN_M6_VGL}Equilibrium simulations at $T=250\,\mathrm{K}$. 
    Shown is the width $l$ of B stripes as a function of $\erg{AB}$
    for the Lennard-Jones and the Morse ($a=6.0$) potential for different values of $\varepsilon$.
    Each data point is obtained by averaging over three independent simulation runs.
  }
\end{figure}

As Fig.\ \ref{LEN_M6_VGL} points out, the main difference one observes is that for 
the same misfit and $\erg{AB}\leq 0.6\erg{A}$ the
stripes for the Morse potential are systematically thicker, whereas at higher values of $\erg{AB}$ 
the mean stripe width is nearly identical for both potentials at a given misfit.
However, even at values $\erg{AB}\leq 0.6\erg{A}$ the deviations are small compared to the 
influence of the particle concentration on the stripe width.

The equilibrium simulations with the off-lattice model
show that the combination of the binding energy $\erg{AB} > 0$ 
between A and B particles together with the misfit $\varepsilon > 0$ yields
regular patterns of alternating stripes.
This morphology is produced for a wide 
range of parameters and independently of the details of the interactions.
The width of the stripes is 
controlled by the value of $\varepsilon$ together with
the binding energy.


\subsection{Island morphology under non-equilibrium conditions}
\label{non-equilibrium}

In the following we will address the question whether the evolution of a
system, which is governed by a competition between strain and binding energy, 
under non-equilibrium growth
conditions yields similar morphologies as the ones observed in thermal equilibrium.
Therefore, we perform kinetic Monte Carlo (KMC) simulations with 
an increasing number of particles.
Two microscopic processes are taken into account: 
(i) random deposition of adsorbate particles, and 
(ii) diffusion of adatoms on the surface. 
Since desorption of adsorbate particles is negligible
in the considered temperature regime, it is not included in our simulations.
Growth takes place on a $100 \times 100$ substrate of six layers height with fixed bottom layer 
and periodic boundary conditions in $x$- and $y$-direction. 
For all simulation runs the deposition rate for both types of
particles is set to $5 \times 10^{-3}\,\mathrm{ML\,s^{-1}}$.
Thus, the resulting overall deposition rate is
$R_d = 10^{-2}\,\mathrm{ML\,s^{-1}}$. 
The simulations are halted when half the substrate is covered with
adsorbate particles.
Since we are only interested in
the submonolayer regime we disregard second layer nucleation, i.e.\ particles which are 
deposited onto other particles will be ignored. 
Jumps of particles onto others are suppressed for the same reason.
The diffusion of adatoms is described by thermally activated hopping processes between neighboring
binding sites with Arrhenius rates \cite{newman:montecarlo:1999}
\begin{equation}\label{arrhenius}
  R = \nu \, \exp\left(-\frac{E_{a}}{k_B T}\right).
\end{equation}
We use $\nu = 10^{12}\,\mathrm{s^{-1}}$ as common attempt frequency for all diffusion events.
The activation energy $E_a$ for a diffusion jump of a particle between two binding sites 
is given by
$E_a = E_t - E_b$ where $E_t$ and $E_b$ are the potential energies of the particle at 
the transition state and the initial binding site, respectively.
Since in the considered misfit regime dislocations do not appear,
$E_b$ can be determined rather easily by placing the particle
on the perfect square lattice site and subsequent relaxation
with respect to the precise, continuous particle positions \cite{much:diss:2003}.
The calculation of $E_t$ implies searching for a first order saddle point in the 
potential energy surface (PES) generated by the superposition of all pair-interactions 
according to Eq.\ (\ref{CUBIC}) \cite{much:diss:2003}. 
This is achieved by an iterative algorithm,
the so-called activation-relaxation technique (ART) \cite{barkema:prl77,mousseau:pre57}.

As interaction strength between A and B particles we choose
\begin{equation}\label{E_AB}
  \erg{AB} = 0.6\,\erg{A},
\end{equation}
which---under equilibrium conditions---leads to the formation of rather thick stripes and 
for which the influence of the misfit should be clearly observable.
On the basis of the equilibrium simulation results, we expect also a noticeable
dependence on the choice of the potential for this interaction strength.
Note that $\erg{A}=\erg{B}$ are given according to Eq.\ (\ref{E_S})
and $\erg{S}$ is specified in Table \ref{TAB_ES} for the different potentials.

\begin{figure}[t]
  \begin{minipage}{0.99 \textwidth}
    \centering
    \epsfxsize= 0.8\textwidth
    \epsffile{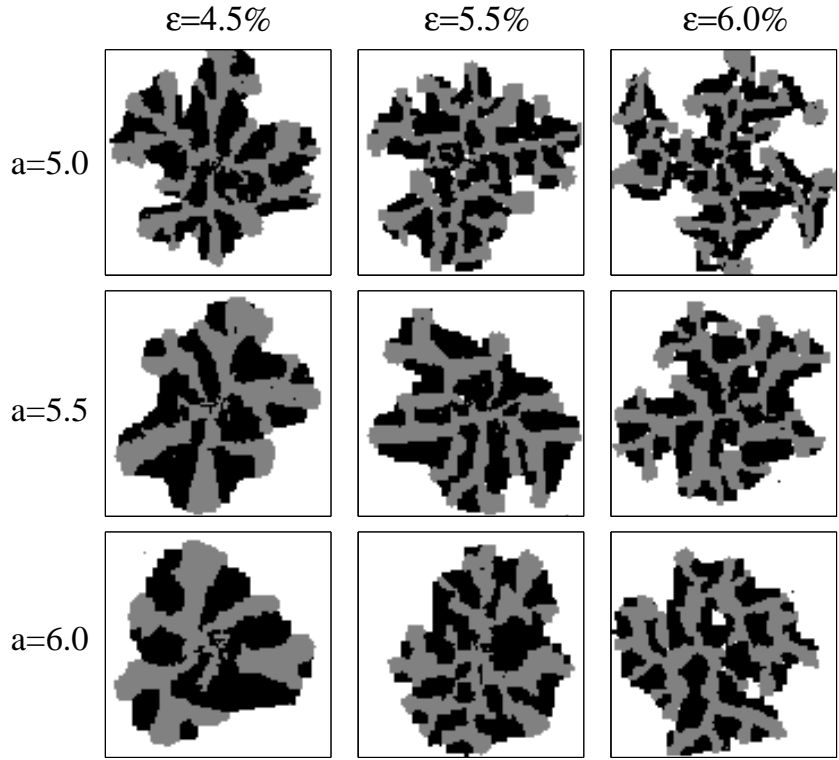}
  \end{minipage}
  \caption{Exemplary surface configurations obtained by KMC simulations with the Morse potential 
    [Eq.\ (\ref{MORSE})]
    for various values of the parameter $a$ and misfit $\varepsilon$.
    The bigger B particles are shown in light gray.}
  \label{OFF_SNAPS}
\end{figure}

This choice of the potential depth yields a higher barrier for edge diffusion
than for diffusion on plain substrate in our simulations.
However, the barrier for edge diffusion is still smaller than that for detachment from the edge.
So particles attached to an island edge are more likely to diffuse there than to detach.
This is of particular importance since we focus here on phenomena, where edge diffusion is
supposed to have a strong impact 
(cf.\ Sec.\ \ref{introduction} and \cite{hwang:prl76,hwang:chemrev97}).
Note also that for the cubic lattice [Eq.\ (\ref{CUBIC})] diagonal diffusion jumps
can be neglected since they imply traversing a maximum in the PES \cite{much:diss:2003}.
The kinetic Monte Carlo simulations are carried out following the standard scheme where in each
Monte Carlo step an event $k$ (deposition or diffusion) is chosen according to its rate $R_k$ 
and performed \cite{newman:montecarlo:1999}.
The crystal is then locally relaxed around the location of the event and the rates for all 
events affected by this relaxation are re-evaluated. 
The system time is incremented by an interval $\tau$ which is chosen
from an exponential distribution \cite{newman:montecarlo:1999}. 
Similar to the equilibrium simulations,
a relaxation of the entire system is performed after
$4 \times 10^5$ steps in order to avoid strain accumulation.

We present now results on the influence of the misfit and the used potential at 
a temperature $T = 500\,\mathrm{K}$.
Comparative simulation runs showed that under the same growth conditions both particle types form
compact, rectangular islands if they are deposited alone onto the substrate. 
We also observed for the B particles with positive misfit that an island which 
becomes larger than a critical island size splits up into smaller islands. 
This can be understood as relaxation of the accumulated compressive strain in the island.
Note that a similar effect is observed experimentally for Cu/Ni(110) where copper islands undergo 
a shape transition when they exceed a critical island size \cite{mueller:prl80}.

\begin{figure}[t]
\begin{minipage}{0.45 \textwidth}
  \epsfxsize= 0.99\textwidth
  \epsffile{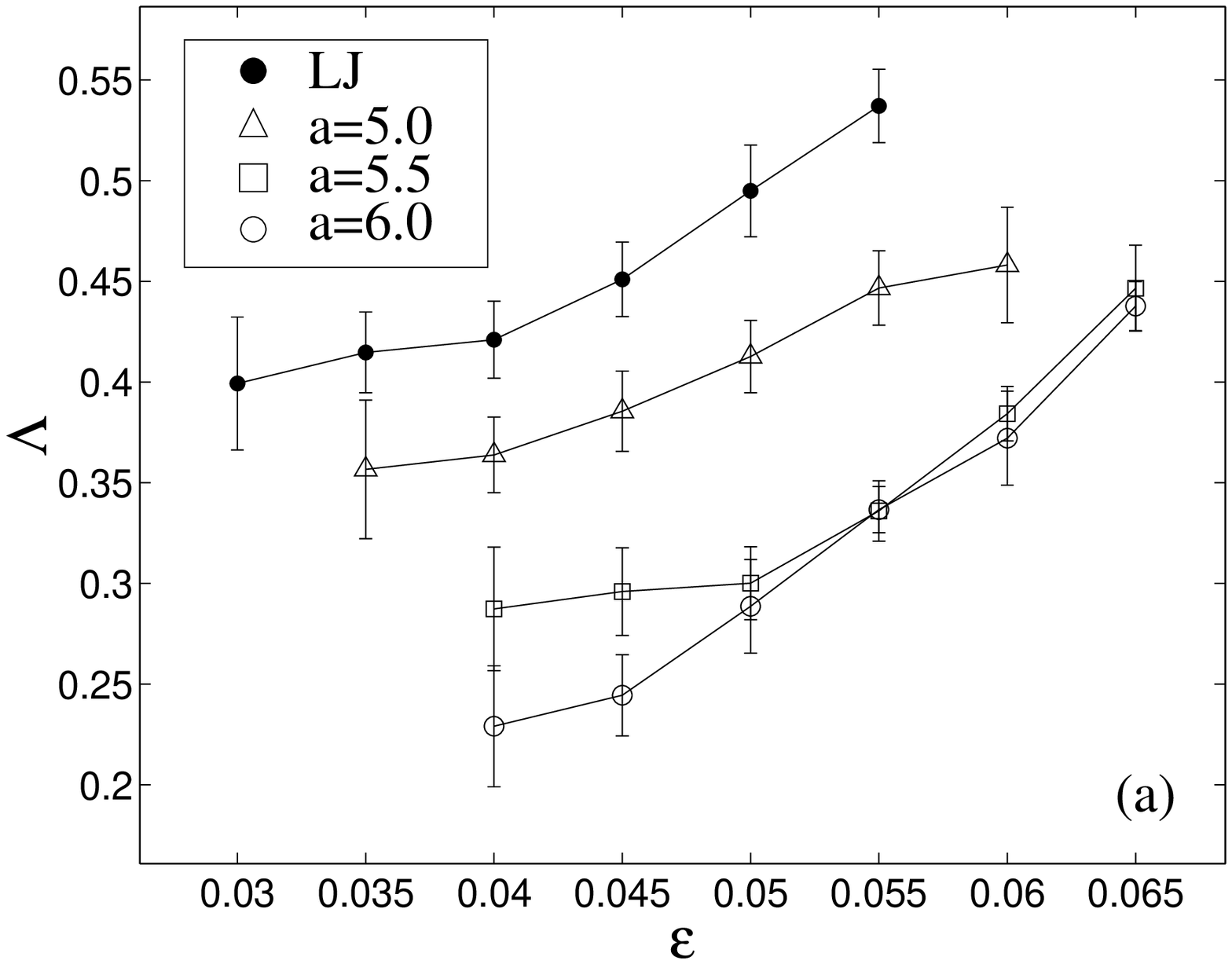}
\end{minipage}
\hfill
\begin{minipage}{0.45 \textwidth}
  \epsfxsize= 0.97\textwidth
  \epsffile{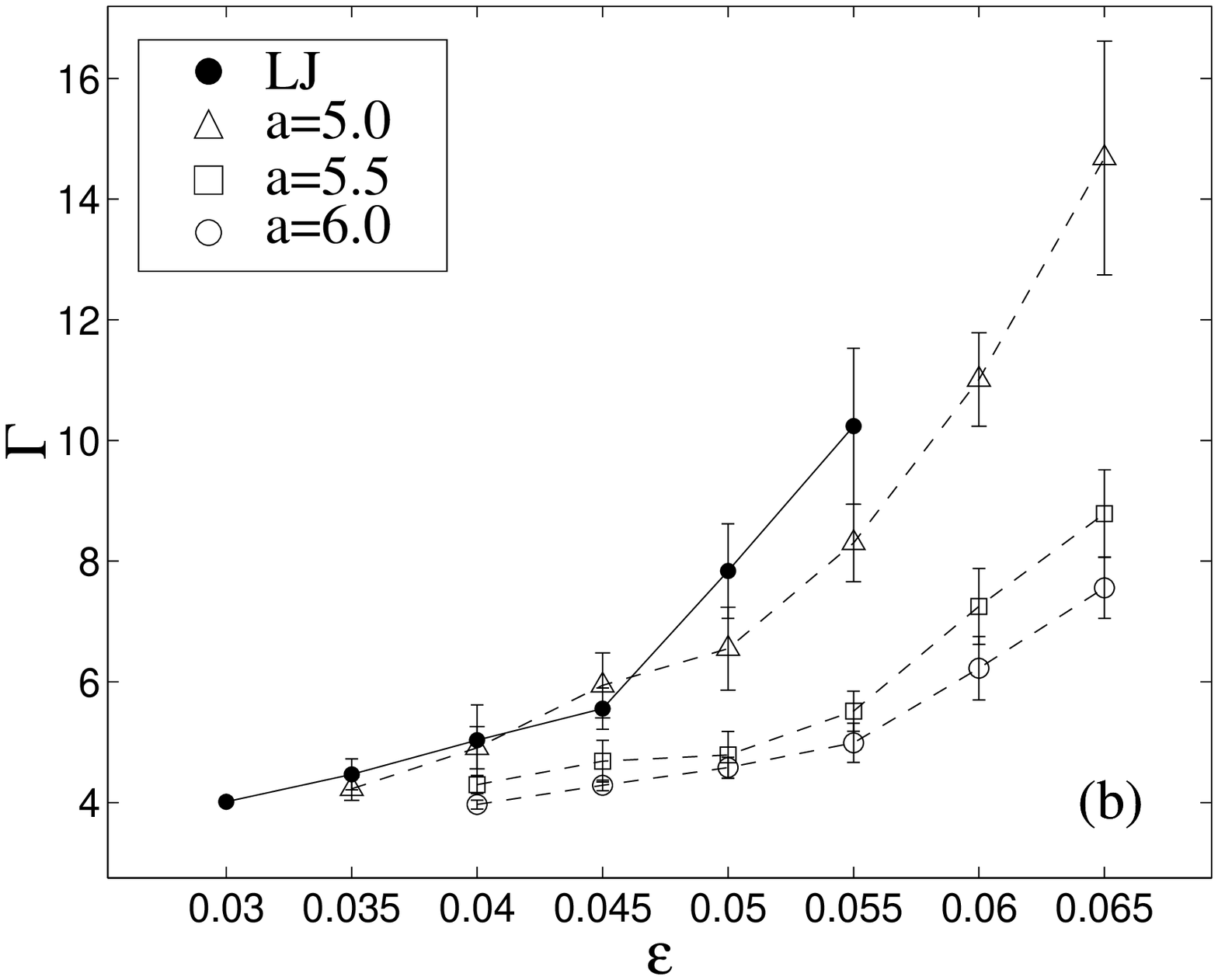}
\end{minipage}
\caption{(a) Ratio $\Lambda$ between perimeter particles and total number of particles in the 
big B clusters
for the used potentials.
(b) The number of perimeter particles divided by the square root of deposited particles $\Gamma$ 
vs. $\varepsilon$.
Each value is obtained by averaging over ten independent simulation runs.
The errorbars are given by the standard deviation.}
\label{QUANT}
\end{figure}

In the case of co-deposition, we observe a completely different situation:
Figure \ref{OFF_SNAPS} shows snapshots of simulation runs for the Morse potential 
[Eq.\ (\ref{MORSE})] for various values of $a$ and $\varepsilon$.
These structures are exemplary for all simulation results:
the B particles (shown in light gray) assemble into a few big clusters. 
With increasing misfit the branches
of these clusters become thinner and of more uniform width.
The A particles surround these branches without showing a similar shape.
It is also seen from Fig.\ \ref{OFF_SNAPS} that with increasing misfit the ramification of 
the structure as a whole increases. 
This is clearly related to the restricted width of the B stripes: a B particle 
rather attaches to the thin end of a stripe.
This implies that thinner stripes of material B (light gray)
grow outwards faster, leading to increasing ramification of the structure.

At a given misfit the B branches are the thinner the smaller the value of $a$ in 
the Morse potential is.
Consequently, at a given misfit the island-ramification is more pronounced for $a=5.0$ than for 
$a=6.0$.
This is in agreement with the equilibrium simulations where a steeper potential yields thicker 
stripes.

In order to quantify the observations we calculate for each connected cluster of B particles
the ratio $\Lambda$ between its perimeter length and its volume.
This is done by counting the
number of perimeter particles together with the total number of particles in the same cluster.
We take only the {\it backbone} of the structures into account and neglect smaller clusters
($<700$ particles).

The ratio $\Lambda$ is a measure for the
average thickness of the cluster, see Fig.\ \ref{QUANT}(a).
For example, for a rather thin cluster most of its particles sit at the
edge and therefore $\Lambda$ should be close to 1, whereas $\Lambda$ should decrease if the 
cluster becomes more compact.
In addition, we measure the species-independent quantity $\Gamma$,
which is given by the number of particles in the system
with less then $4$ nearest neighbors, divided by the square root of
the total number of adatoms.
$\Gamma$ provides a measure for the length of the structure's perimeter and therefore the 
ramification, see Fig.\ \ref{QUANT}(b).
A single perfect quadratic island on the substrate corresponds to $\Gamma \approx 4$,
whereas larger values of $\Gamma$ indicate roughening of the island shape.
The correlation between $\Lambda$ and $\Gamma$ is clearly observable for all used potentials:
$\Lambda$ increases with increasing misfit indicating thinner B clusters. Simultaneously 
the ramification
increases. The formation of B branches of well-defined thickness is a common
phenomenon for the used pair-potentials.

\begin{figure}[t]
  \begin{minipage}{0.45 \textwidth}
    \epsfxsize= 0.99\textwidth
    \epsffile{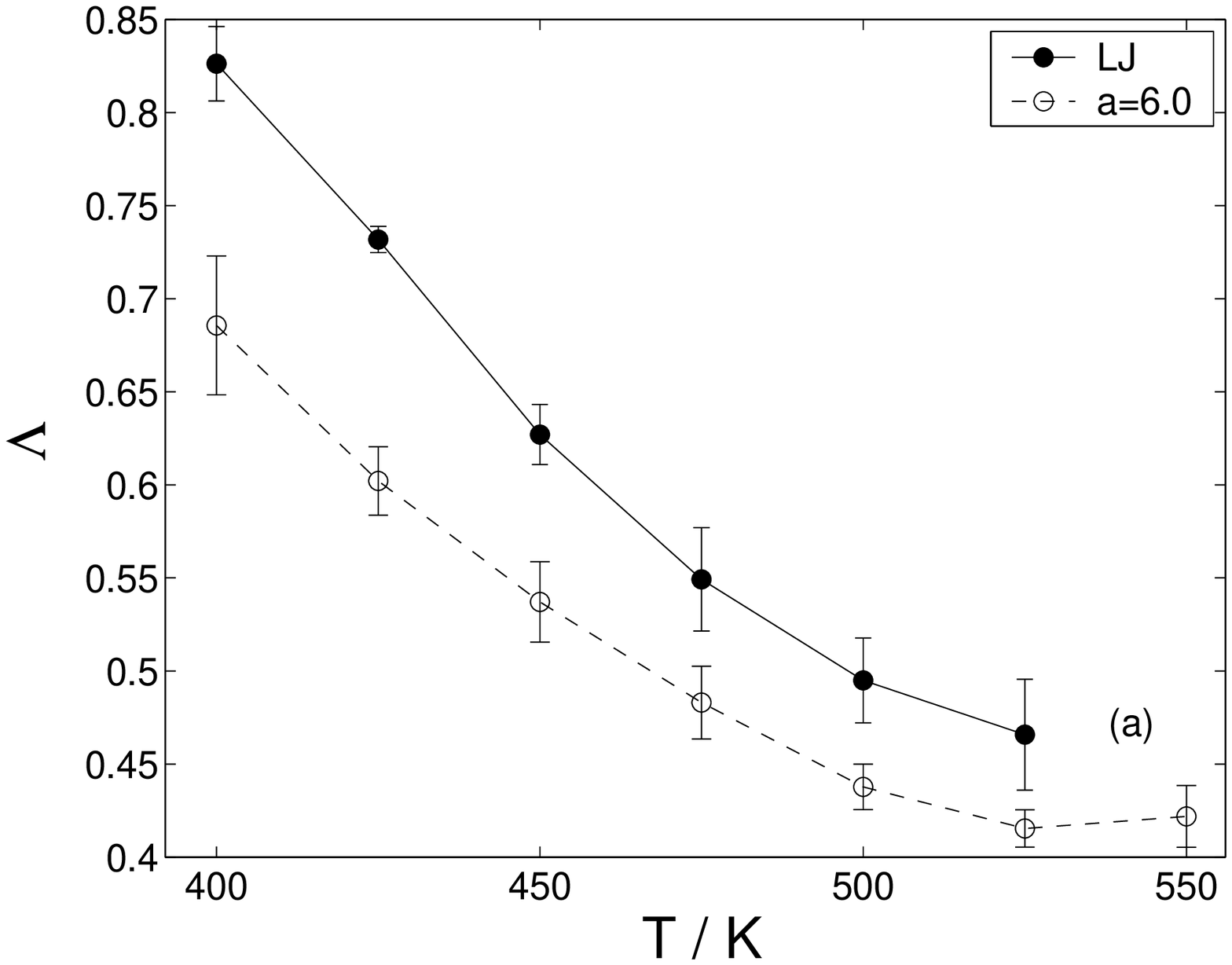}
  \end{minipage}
  \hfill
  \begin{minipage}{0.45 \textwidth}
    \epsfxsize= 0.97\textwidth
    \epsffile{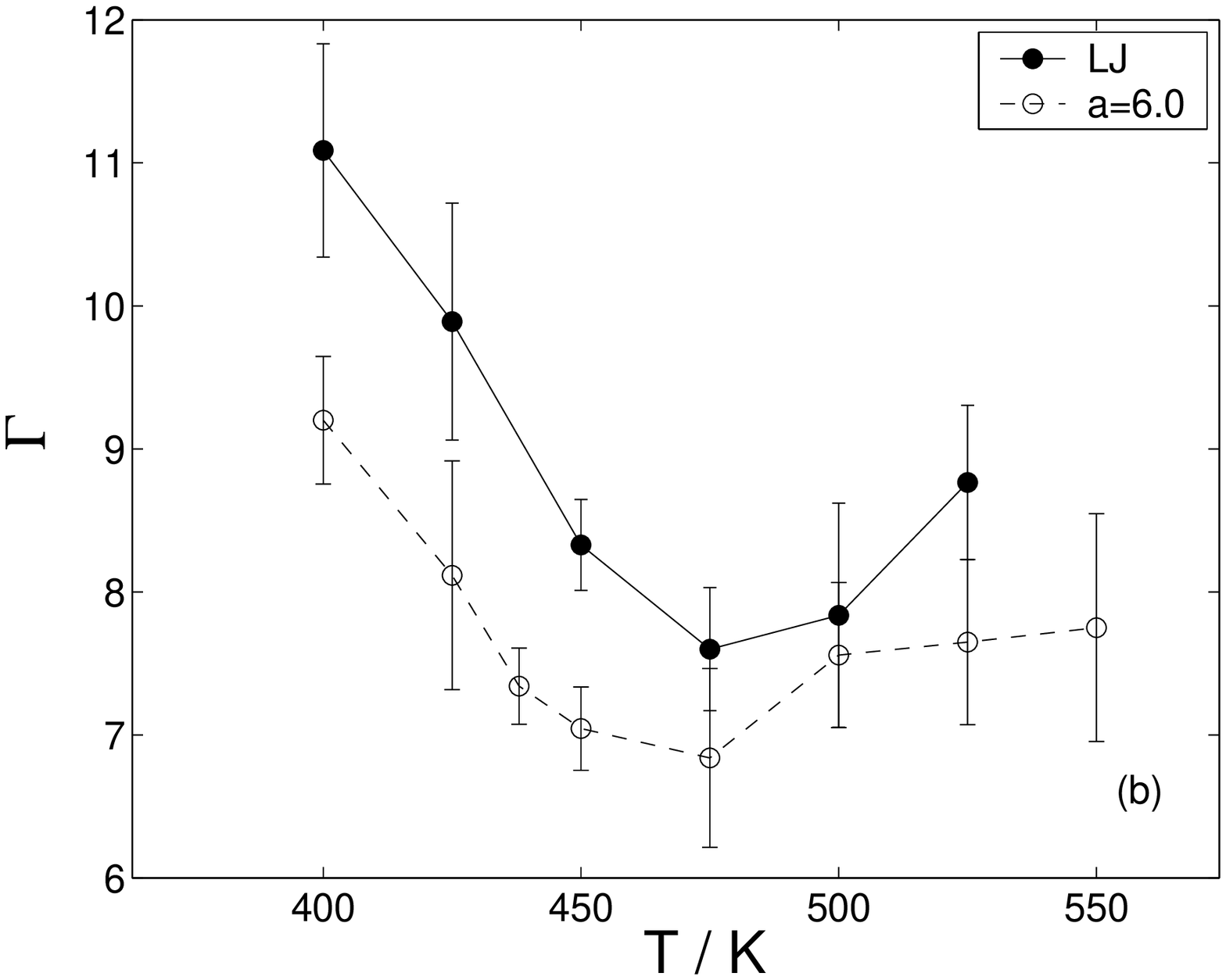}
  \end{minipage}
  \caption{\label{LGvsT}Temperature dependence of (a) $\Lambda$ and (b) $\Gamma$ for 
    the Lennard-Jones potential
    with $\varepsilon=5.0\%$ (filled circles) and the Morse $a=6.0$ potential 
    with $\varepsilon=6.5\%$ (open circles).}
\end{figure}

One might suspect
that the observed ramification of the islands is due to temperature effects, only:
i.e.\ the used temperature may be high enough for the formation of cubic clusters
of a single species, but enlarged edge diffusion barriers in the
case of mixed deposition
might cause dendritic growth at the same temperature.

In order to investigate the temperature dependence of the island morphologies we 
performed simulations for temperatures between $400\,\mathrm{K}$ and $550\,\mathrm{K}$ using
the LJ potential with $\varepsilon=5.0\%$ and 
the Morse potential with  $a=6.0$ and $\varepsilon=6.5\%$.
For the given parameters, strongly ramified islands grow at $T=500\,\mathrm{K}$.
At low temperatures we observe multiple islands due to the reduced diffusion length.
They exhibit frayed edges and rather thin and disordered B stripes.
With increasing temperature the B stripes become wider and
more regular in shape, the island edges become smoother.
The observations are reflected in the temperature dependence of $\Lambda$ and $\Gamma$
as shown in Figs.\ \ref{LGvsT}(a) and (b).
We stress that the ramification $\Gamma$ does not decrease monotonously
with increasing temperature (as one might expect).
For both
potentials it exhibits a minimum at $T\approx 475\,\mathrm{K}$ and then
slowly increases with $T$ for higher temperatures.
This observation clearly rules out that the observed ramification is merely an artefact
of the low growth temperature.

The enhanced mobility of the particles causes a more distinct separation of
the two particle types, resulting in more regular B stripes.
As Fig.\ \ref{LGvsT}(a) shows the width of the B stripes approaches 
a constant value for the high temperature region.
Furthermore, we observed that for high enough temperatures
nearly all B clusters are aligned in the $\left< 11 \right>$ directions in order to
achieve the energetically most favorable arrangement of 
particles like in the equilibrium simulations (see \cite{much:diss:2003}).

\begin{figure}[t]
  \begin{minipage}{0.45 \textwidth}
    \epsfxsize= 0.99\textwidth
    \epsffile{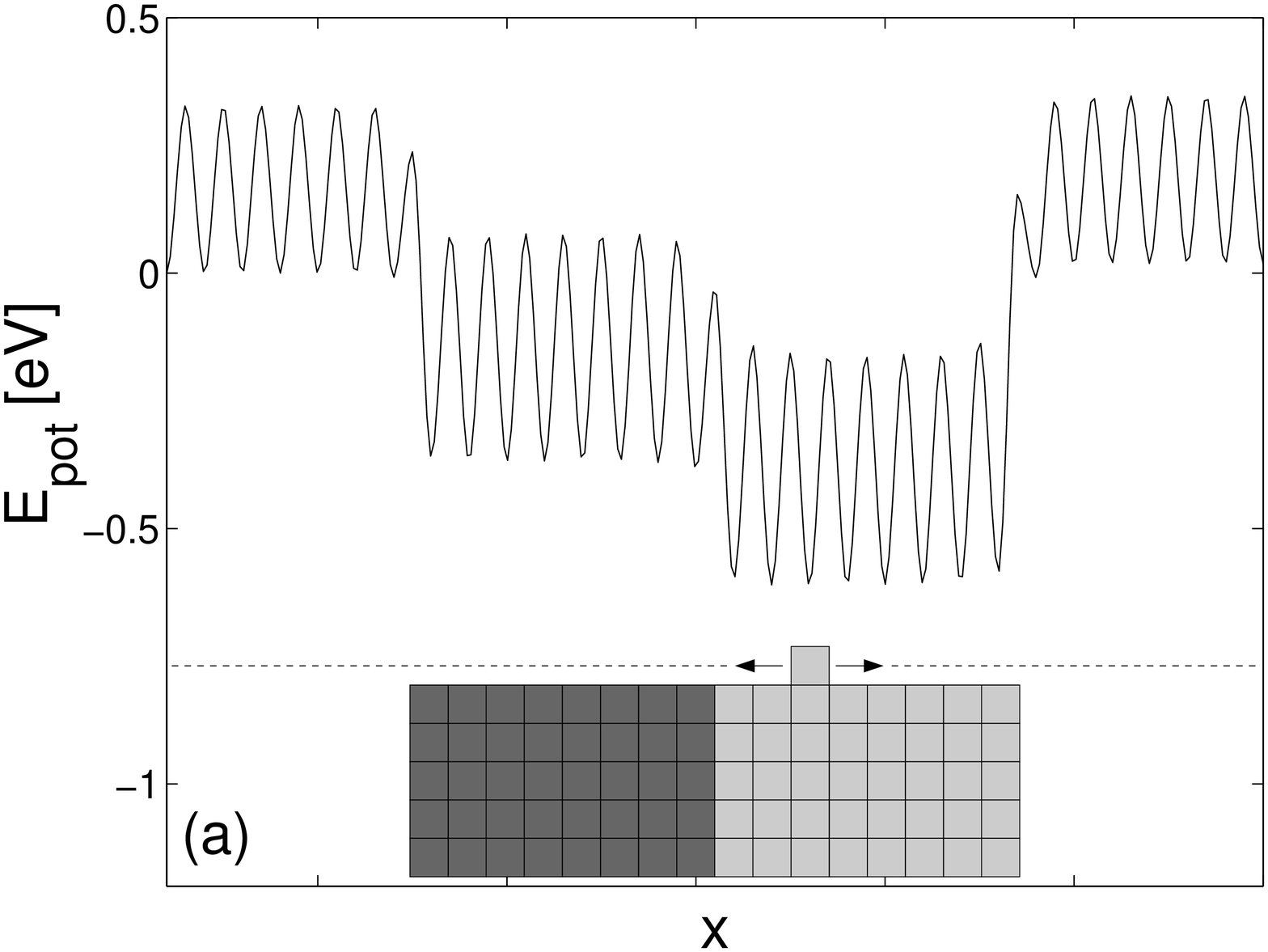}
  \end{minipage}
  \hfill
  \begin{minipage}{0.45 \textwidth}
    \epsfxsize= 0.97\textwidth
    \epsffile{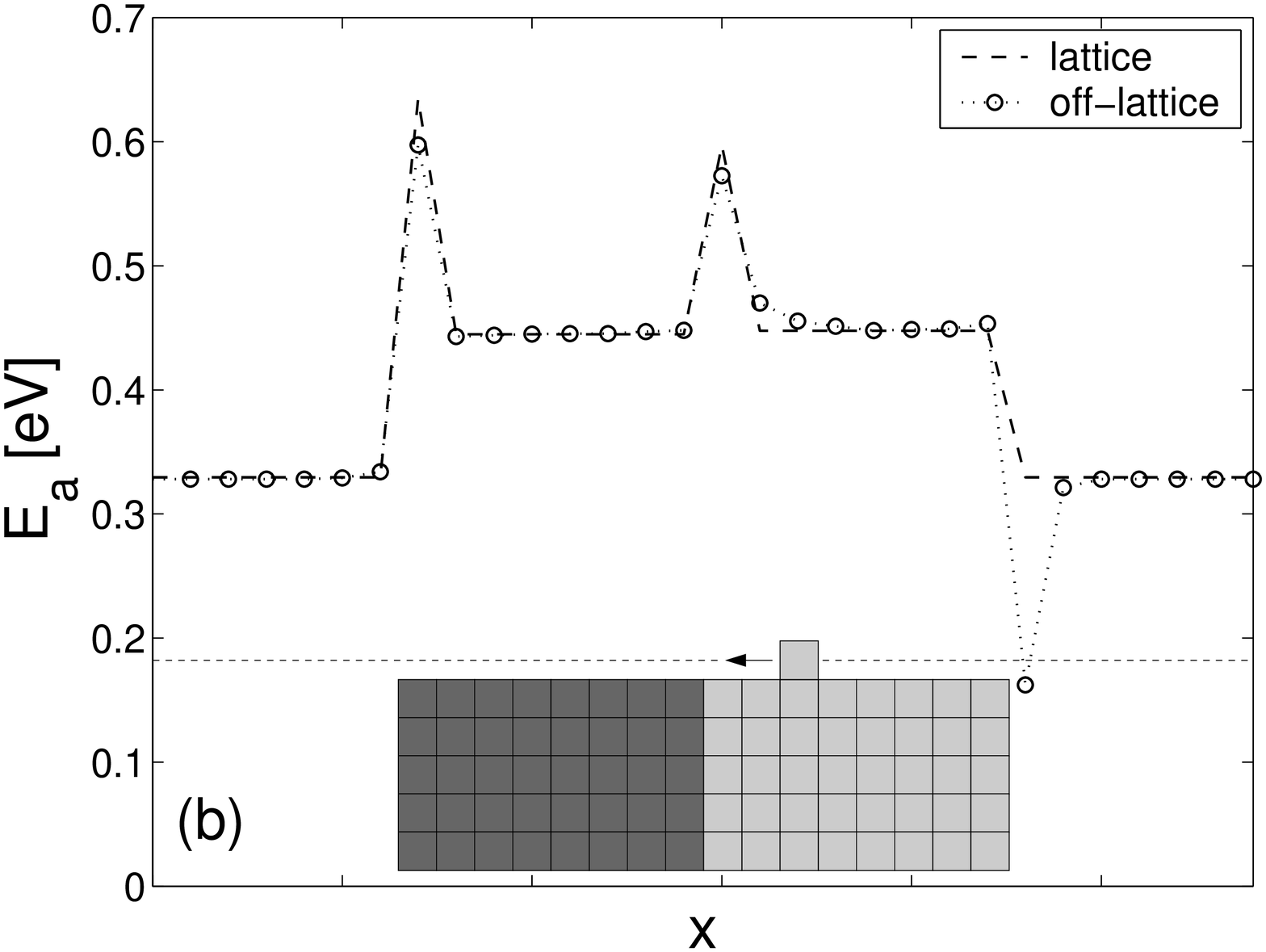}
  \end{minipage}
  \caption{\label{RGOTV}(a) Sketch of the potential energy for a B particle (light gray) 
    diffusing along the dashed line parallel to the step edge of the small A--B cluster.
    The values correspond to the Lennard-Jones potential with 
    $\varepsilon = 4\%$.
    (b) Diffusion barriers for leftward jumps of the B particle from (a). The panel shows 
    values obtained
    from the off-lattice model (open circles) and as used in the lattice gas approximation 
    (dashed line) discussed in
    Sec.\ \ref{lattice-gas-model}.}
\end{figure}

The question now is in which way the observed branches are related to the stripe structures
found in the equilibrium simulations.
Figure \ref{RGOTV}(a) shows the potential energy for a B particle diffusing near an A--B interface
for $\varepsilon=4\%$ and $\erg{AB}=0.6\erg{A}$.
The weaker A--B interaction causes an extra step edge diffusion barrier for the jump from
the B to the A region. This can be more clearly seen in Fig.\ \ref{RGOTV}(b) where 
the diffusion barrier for a jump
to the left is plotted versus the particle position. The diffusion barriers are given by 
the energy difference
between the corresponding transition state energy and the binding energy. A similar plot 
is obtained for the
rightward diffusion jumps of an A particle.
As already mentioned in Sec.\ \ref{introduction}
the enhanced diffusion barrier at the A--B interface is believed to favor the formation of 
alternating stripes.

In the following section, we discuss, by means of a lattice gas model,
how such a diffusion barrier influences the multi-component growth.
Of special interest is the question, whether
the stripe formation and the island morphology, as observed in our
off-lattice simulations, can be explained in a simplifying framework.
Clearly, strain effects cannot be taken into account explicitly
in a pre-defined lattice of possible adatom sites with fixed distances.


\section{Lattice description}
\label{lattice-description}

\subsection{Lattice gas simulation model}
\label{lattice-gas-model}

In our lattice gas model, two adsorbate species A and B grow on a square substrate S
with $150 \times 150$ adsorption sites.
Unlike the off-lattice model where a particle interacts with all particles within the
range of the potential, A and B particles interact now only
with their lateral nearest neighbors through attractive two-particle
interactions with the energy parameters
$\Erg{AA}$, $\Erg{BB}$ and $\Erg{AB}$. Here,
$\Erg{AA}$, $\Erg{BB}$ denote the binding of two A-particles or two B-particles, respectively, and 
$\Erg{AB}$ represents the interaction of an A-particle with a B-particle. 
The total energy of the system can then be written as
\begin{equation}\label{latticegas_hamiltonian}
  H = -\Erg{AA} \num{AA} - \Erg{BB} \num{BB} - \Erg{AB} \num{AB} + \pot{A} \num{A} + 
  \pot{B} \num{B} 
\end{equation}
where $\num{A}, \num{B}$ denote the number of A and B particles, 
and $\num{AA}, \num{BB}, \num{AB}$ 
count the number of A--A, B--B and A--B bonds, respectively.
The binding of adsorbate particles to the
substrate is represented by the effective chemical potentials $\pot{A}$ and $\pot{B}$.
Diffusion of adatoms on the surface is described by thermally activated 
nearest-neighbor hopping processes with Arrhenius rates 
$R_i=\nu \exp(-E_{a,i}/k_B T)$, 
where we use again 
$\nu=10^{12}\,\mathrm{s^{-1}}$ as common attempt frequency.
The temperature $T$ is set to $500\,\mathrm{K}$.

A diffusion event $i$ which leads from the starting (s) to the
final (f) configuration 
is modeled using Kawasaki type energy barriers \cite{newman:montecarlo:1999}
with the activation energy 
\begin{equation}
  \label{kawasaki_barriers}
  E_{a,i} = \max \{B_{s,i}, B_{f,i} + \Delta H_i\}.
\end{equation}
Here, $\Delta H_i$ denotes the total energy change caused by the diffusion event $i$ 
which in turn is given by Eq.\ (\ref{latticegas_hamiltonian}).
In general, the diffusion barriers $B_{s,i}$ and $B_{f,i}$ 
may depend on the type of the diffusing particle as well as 
the starting and final configuration of the system.

In order to obtain a general insight into the behavior of the model,
we consider first a simplified version of our model where A and B particles are treated in 
a symmetric way, 
and where the inter-species binding energy is the key characteristics. 
In Sec.\ \ref{modified-model},
we will use a modified parameter set 
in order to achieve a comparison between lattice and off-lattice simulations.


\subsection{Symmetric treatment of adsorbate species}
\label{simplified-model}

In this section, 
we assume that 
all barriers $B_{s,i}$, $B_{f,i}$ in Eq.\ (\ref{kawasaki_barriers}) are equal, i.e.\ 
$B_{s,i} = B_{f,i} = B_0$ for all $i$. 
Also, the strength of A--A and B--B bonds will be the same: 
$\Erg{AA} = \Erg{BB} = E^0$ .
The model is 
governed by the interaction $\Erg{AB}$ between A and B particles 
which is assumed to be weaker than between two particles of the same type: 
$\Erg{AB} < E^0$, 
following the hypothesis in \cite{hwang:prl76}. 
This has two main implications for a particle diffusing
along the step edge of an A--B cluster as the one depicted in Fig.\ \ref{RGOTV}. 
First, the particle is facing an enhanced diffusion barrier when crossing a domain wall.
For example, a B particle faces a diffusion barrier $B^0 + E^0 - \Erg{AB} > B^0$ 
when it attempts to cross the A--B interface coming from the B side 
where it is bound more strongly (note, that for the reverse jump, the barrier is $B^0$), 
cf.\ Fig \ref{RGOTV}(b).
The same happens to an A particle which tries to cross the interface coming from the A side. 
Thus, A and B particles diffusing along step edges are likely to be reflected at A--B interfaces. 
Second, the activation energy for detachment 
of a particle of A or B type from a step edge made up of the opposite type is lower 
than that for detachment from a step edge of the same type. 
The two effects combined reflect basically the influence of a weaker A--B interaction 
in the off-lattice simulations, disregarding though all influences of strain or long 
range interactions.

To investigate the influence of the binding energy $\Erg{AB}$, 
we fix $B^0 = 0.37\,\mathrm{eV}$ and $E^0 = 0.51\,\mathrm{eV}$. 
This reproduces roughly the homoepitaxy
($\varepsilon = 0$) barriers for
diffusion on planar substrate and detachment from an island edge as measured in 
the off-lattice simulations.
$\Erg{AB}$ is varied between $0.31E^0$ and $0.71E^0$.
Following the off-lattice simulation, in all simulation runs the deposition rate for 
both types of particles is set to
$5\times 10^{-3} \,\mathrm{ML\,s^{-1}}$ resulting in an overall deposition rate of 
$R_d=10^{-2}\,\mathrm{ML\,s^{-1}}$.
When the total adsorbate coverage has reached $0.5\,\mathrm{ML}$ the simulation is halted.

\begin{figure}[t]
  \begin{minipage}{0.24 \textwidth}
    \epsfxsize= 0.99\textwidth
    \epsffile{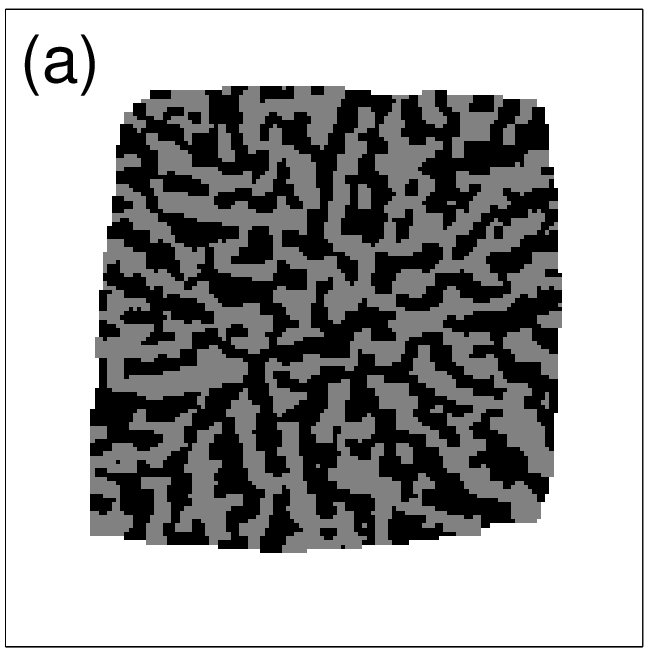}
  \end{minipage}
  \hfill
  \begin{minipage}{0.24 \textwidth}
    \epsfxsize= 0.99\textwidth
    \epsffile{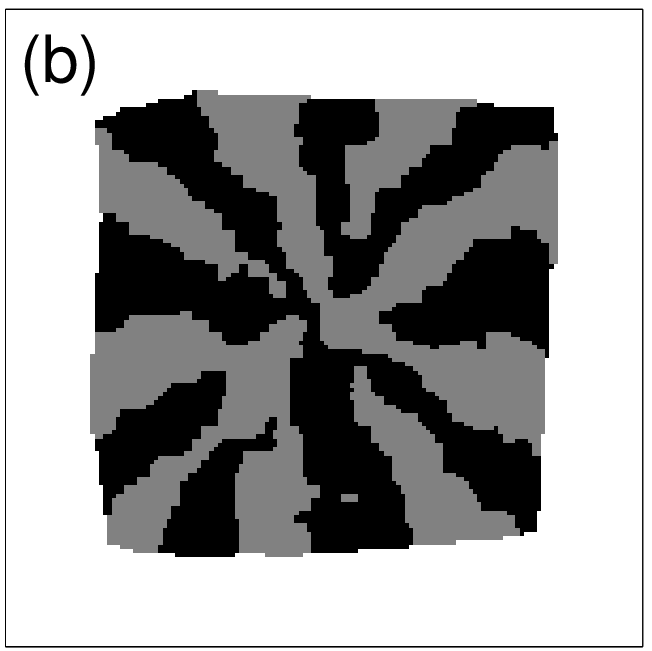}
  \end{minipage}
  \hfill
  \begin{minipage}{0.24 \textwidth}
    \epsfxsize= 0.99\textwidth
    \epsffile{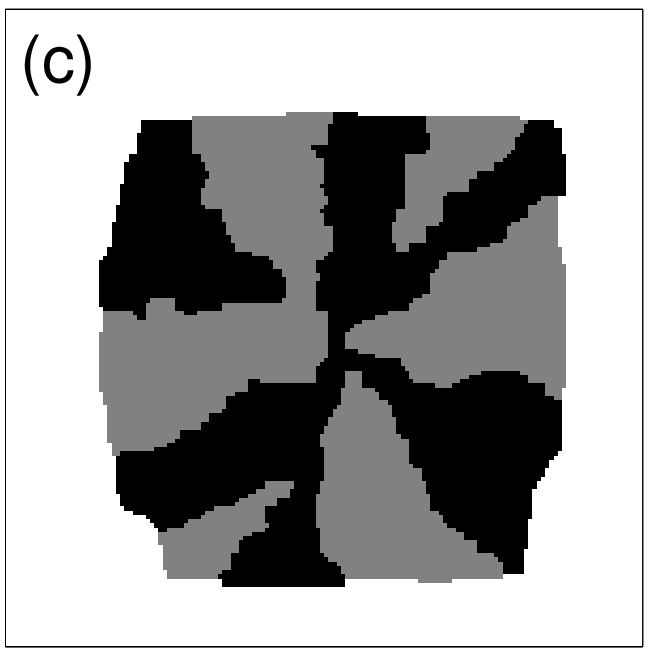}
  \end{minipage}
   \hfill
   \begin{minipage}{0.24\textwidth}
    \epsfxsize= 0.99\textwidth
    \epsffile{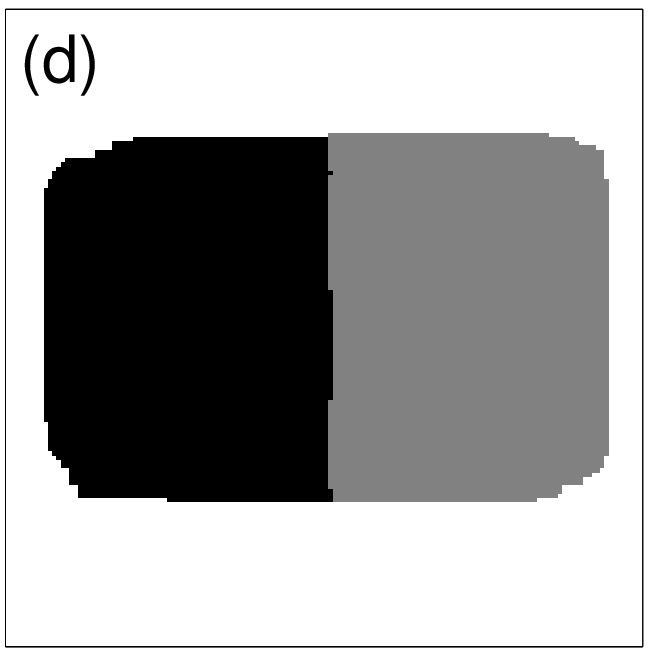}
  \end{minipage}
  \caption{\label{LATI}Lattice gas simulations with symmetric treatment of A and B particles.
    Island configurations obtained under non-equilibrium growth with $T=500\,\mathrm{K}$ and 
    $E^0=0.51\,\mathrm{eV}$ for
    $\Erg{AB}=0.71E^0, 0.51 E^0, 0.47E^0$ (a)--(c). 
    (d) System configuration obtained after
    $t = 3 \times 10^4\,\mathrm{s}$ of equilibrium simulation with $T=500\,\mathrm{K}$, 
    $E^0=0.51\,\mathrm{eV}$ and 
    $\Erg{AB} = 0.26\,\mathrm{eV}$. The system size
    is $150 \times 150$ and the total coverage is 
    $\Theta=0.5\,\mathrm{ML}$.}
\end{figure}

Figure \ref{LATI}(a)--(c) shows exemplary configurations obtained 
at the end of simulation runs
for different values of the binding energy $\Erg{AB}$.
For all values of $\Erg{AB}$ one observes compact island shapes with the island boundaries 
roughly parallel to the lattice directions. 
The weaker binding energy between A and B particles leads to 
an aggregation of
particles of the same type in clusters which can be characterized as stripes.
While for the higher value of $\Erg{AB}$ these stripes are rather thin
and show a considerable degree of irregular intermixing for lower values of $\Erg{AB}$ the
stripes are both much thicker and there is a tendency for them to stretch outwards. 
One also sees that at
a certain stage of the island growth a stripe of one particle type may become wide enough 
for particles of
the other type to form a stable nucleus within this stripe, thus leading to 
a branch-like structure.
Similar interplay between growth kinetics and phase ordering has been observed
in a simple model with line geometry \cite{kotrla:epl39}.

The occurrence of the stripe-like structures and the branching under non-equilibrium conditions 
must be attributed to the kinetic
segregation of A and B particles. From thermodynamic considerations one expects more or 
less complete separation of 
both particle types for not too high temperatures and not too large values of $\Erg{AB}$. 
We have tested this assumption by performing canonical equilibrium simulations 
with fixed adsorbate coverages $\num{A}/n = \num{B}/n = 0.25$ ($n=\num{A}+\num{B}$) 
and a random distribution of particles as initial condition. 
Similar to the off-lattice equilibrium simulations (Sec.\ \ref{equilibrium})
we apply a non-local dynamics where in each step an A or B particle
from site $i$ may jump to {\em any\/} vacant lattice site $j$.

Figure \ref{LATI}(d) shows a typical system configuration 
for $\Erg{AB} = 0.26\,\mathrm{eV}$ and $T = 500\,\mathrm{K}$ obtained 
after $3\times 10^4\,\mathrm{s}$ simulated time which confirms that A and B particles separate 
and due to the attractive binding energy $\Erg{AB}$ 
form a single rectangular island consisting of one A and one B region. 
Note that the interface between the A and the B region is not perfectly straight and the 
island edges are rounded, in accordance with theoretical calculations which yield $T_R=0$ as 
roughening temperature of two-dimensional crystals 
\cite{pimpinelli:growth:1998}.
Similar results are obtained for various values of $\Erg{AB}$ and temperature $T$.

We conclude from our lattice gas simulations, that the step edge barrier indeed gives 
reason for stripe formation. 
The equilibrium simulations show that the formation of
stripes can be traced back to the kinetic segregation of A and B particles under non-equilibrium
growth conditions. 
The width of the 
stripes can be controlled by adjusting the
binding energy between A and B particles. However as Fig.\ \ref{LATI} shows neither 
asymmetries between A and B clusters nor ramification of the islands is observed here.
This is not surprising since A and B particles were treated in a
symmetric way, whereas in the off-lattice simulations the different sign of the misfits causes
different diffusion barriers for A and B particles. 
For example, 
the substrate diffusion of the B particles with positive misfit is always faster than that of 
the A particles with negative misfit \cite{much:epl63,much:diss:2003,schroeder:surf375}.
Furthermore in the off-lattice method the barriers for edge diffusion are higher than 
the substrate diffusion barriers. This could also give rise to a ramified island morphology.


\subsection{Comparison of lattice and off-lattice formulation}
\label{modified-model}

To account for basic differences of the two particle types in our lattice gas model we now 
use a modified parameter set
which is fitted to reproduce the barriers of characteristic diffusion processes in 
the off-lattice model.
The question is whether a simple misfit dependence of the diffusion barriers could
lead to the observed results (e.g.\ island ramification) within such an enhanced 
lattice gas model.

Therefore, we extract the barriers for free diffusion on the substrate
as well as averaged values for edge diffusion and detachment for a {\it fixed} island size
(see also Fig.\ \ref{RGOTV}) as a function of the misfit. These barriers are then used to determine
$\Erg{AA}$, $\Erg{BB}$ and $\Erg{AB}$ as well as the $B_{s,i}$ and $B_{f,i}$ for the different 
diffusion processes
[cf.\ Eq.\ (\ref{kawasaki_barriers})].
Thus, the modified lattice model incorporates the basic misfit dependence of the diffusion 
barriers.
However, effects of the long range interaction, like the reduced barrier for jumps towards 
an island (cf.\ Fig.\ \ref{RGOTV}) still have to be neglected here.

\begin{figure}[t]
  \begin{minipage}{0.99 \textwidth}
    \epsfxsize= 0.99\textwidth
    \epsffile{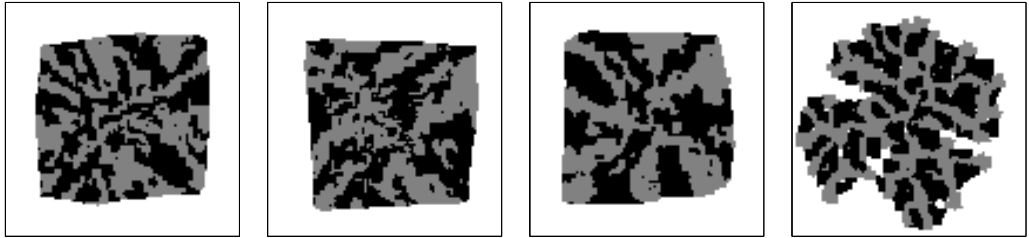}
  \end{minipage}
  \caption{\label{COMP_O_L}Comparison of snapshots  for the enhanced lattice and 
    the off-lattice model in the case of the Lennard-Jones potential.
    The panels show (from left to right) lattice/off-lattice results for $\varepsilon=0$ 
    and lattice/off-lattice results for  $\varepsilon=5\%$. B particles appear in light gray.}
\end{figure}

\begin{figure}[t]
\begin{minipage}{0.45 \textwidth}
  \epsfxsize= 0.99\textwidth
  \epsffile{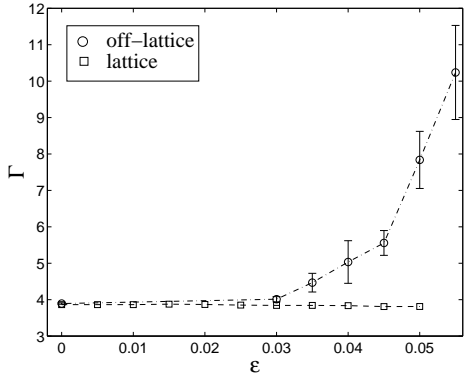}
\end{minipage}
\caption{\label{RAMCOMP}Island-ramification for lattice and off-lattice simulations.
Each value is obtained by averaging over ten independent simulation runs.
For the off-lattice simulations errorbars are given by the standard deviation. 
For the lattice simulations errorbars are smaller than the symbols.}
\end{figure}

Figure \ref{COMP_O_L} shows a comparison between the lattice model and the off-lattice 
simulation for the LJ potential. 
Similar results are obtained by fitting the parameters of the lattice gas model
to the barriers obtained for the Morse potential.
As expected, the islands for both models look very much alike in the case of zero misfit.
However, for $\varepsilon=5\%$ lattice and off-lattice results seem to have little in common.
In the case of the lattice model, 
the separation of A and B regions is more pronounced as for $\varepsilon=0$ 
but neither size limitation of the stripes 
nor island ramification is observable here.
On the other hand, asymmetry of the particle species and island ramification are clearly 
noticeable in the off-lattice configurations.
To quantify our observations
we have measured the ramification $\Gamma$ for both lattice and off-lattice
simulation results.
Figure \ref{RAMCOMP} shows $\Gamma$ for various values of the misfit $\varepsilon$. 
For $\varepsilon = 0$ the islands are roughly quadratic in both types of simulations 
and thus the curves coincide at $\Gamma \approx 4$. 
With increasing misfit the islands in the off-lattice simulations become more and more ramified 
leading to a significant increase of $\Gamma$ for $\varepsilon > 3\%$. 
For the lattice simulations though $\Gamma$ remains constant, i.e.\ no ramification is observed. 

From additional off-lattice simulations, where the reduced barrier for jumps towards an island
is suppressed we find that 
the resulting islands are less ramified whereas the width of the B branches remains
unchanged \cite{much:diss:2003}. 
The reduced island ramification can be traced back to a higher mobility of the particles:
once a particle detaches from an island it has the same probability for jumps towards 
the island as away from it.
The capturing of diffusing adatoms by islands is therefore less pronounced and  
the particles are more uniformly distributed around the island \cite{much:diss:2003}.

These examinations clearly demonstrate that species-dependent diffusion barriers at edges alone
are not sufficient to explain the width restriction of the B branches or the 
ramification of the islands with increasing misfit. Actually, further non-local effects like e.g.\
the above mentioned adatom capturing play a decisive role.
Our enhanced lattice gas model with fitted diffusion barriers thus lacks important
features observed in both experiment and off-lattice simulations.


\section{Summary and Discussion}
\label{summary}

We have studied two-component pattern formation and island shape ramification in a
ternary material system: an adlayer composed of two immiscible components A and B 
deposited on a substrate S of intermediate lattice spacing. 
We have developed and studied atomistic models in order to
investigate different mechanisms of pattern formation
suggested in the literature on the atomistic level.
We have compared results obtained with an off-lattice model (with different interaction
potentials), and a lattice gas model (with different parameterizations). 
In all considered models, the inter-species binding is weaker than the binding between
species of the same kind. In the off-lattice case the explicit incorporation 
of adsorbate misfits is possible whereas in the 
lattice gas description,
this
feature can be taken into account only indirectly by a modification of the parameters. 
The combination of both types of models has enabled us to assess the role of the two main
mechanisms considered as the driving force of stripe formation:
strain relaxation, and kinetic segregation of the elements.
We have performed both equilibrium and non-equilibrium 
simulations for the two different
model types.

Equilibrium simulations using the off-lattice model have been performed for
a completely filled monolayer. 
The results have shown that the adsorbate materials 
segregate and form nanoscale stripes with straight boundaries and a stable well
defined width. 
The stripe width decreases with increasing relative misfits and with
increasing inter-species binding energy. 
Our results indicate also that the stripe width changes with 
the concentration. 
The B particles (positive misfit) form thinner stripes at high B concentration than 
the A particles (negative misfit) at high A concentration. 
We have observed very similar behavior for
different pair-potentials (Lennard-Jones and Morse potential).

The situation is different
in the case of equilibrium simulations using the lattice gas model.
Here, the system undergoes a complete phase separation with a temperature
dependent time of separation. 
Hence, no stripe structure is formed in the long time limit.

Under non-equilibrium conditions, we have studied the growth of an isolated island. 
In the off-lattice simulations
we have observed the formation of highly ramified monolayer islands
with a vein structure similar to that observed in experiments \cite{hwang:prl76}.
A pronounced asymmetry is found in the sense that the bigger B
particles form a backbone of ramified branches,
with the smaller A particles filling in the gaps.
While the shape of 
mixed A--B islands
is ramified, we have observed 
that islands composed of only 
A or B particles
have regular square shape.
This agrees with experimental observations \cite{hwang:prl76}. 
The island ramification has been observed
for different interaction potentials. 
Our results indicate that the ramification 
of two-component islands is
not an artefact of low temperature but the result of chemically 
induced step edge barriers in combination with the effects of strain.

We have studied quantitatively the 
dependence of island shape and composition
on the misfit and on the temperature. 
The increase of the misfit causes an increasing ramification, and
the increase of the temperature yields wider and more regular stripes.
Our results confirm that
there is a correlation between the width of the stripes and the smoothness of
island edges. 

With the help of a simple version of the lattice model in which 
the inter-species binding energy
$\Erg{AB}$ is the only relevant parameter,
we have demonstrated that a chemically induced step
edge diffusion barrier is sufficient to cause the formation of 
structures with alternating stripes.
Here, the stripe formation is a purely kinetic effect.
The stripe width is selected by a balance of different kinetic rates and can be
tuned by $\Erg{AB}$.
We have observed that
the interface of the stripes is rather rough.
Moreover, the observed island shapes are regular in contrast to the simultaneous
observation of vein structures and dendritic growth in experiment 
\cite{hwang:prl76} and our off-lattice simulations. 

In order to rule out the possibility that the absence of island ramification 
in the lattice model is merely caused by the oversimplified symmetric 
treatment of A and B particles 
we have constructed and studied a modified lattice model in which
we have tried to represent
the features, i.e.\ energy barriers, of the off-lattice model as faithful
as possible. Nevertheless, the modified model also fails to reproduce the 
ramification observed in the off-lattice simulations.

A more successful lattice based simulation would have to incorporate 
non-local effects. Diffusion barriers can depend on quite large neighborhoods
in the off-lattice model. 
For instance, barriers for diffusion along an island edge should
depend explicitly on the island size and composition. 
The above mentioned
breaking up of pure B clusters at
a characteristic size indicates that the misfit yields island size
dependent barriers for attachment or detachment. Such
effective long-range interactions can be mediated through
elastic deformation of the
substrate, for instance.  Clearly, an explicit incorporation of cluster size
dependent barriers is beyond the scope of a simple lattice gas model and would
destroy its conceptional advantages.  Alternative routes, e.g.\ the evaluation
of the strain energy for a given lattice configuration, have been suggested
and used in the literature, see for instance \cite{meixner:prl:2001}.

For the sake of computational benefits the off-lattice model was formulated for 
the simple cubic lattice and simple pair-potentials. 
Nevertheless, we believe that these simplifications do not affect our qualitative conclusions.
The model can be modified and extended in different ways to study specific questions.
In order to obtain a closer comparison with experiments on fcc(111) or
fcc(110) surfaces one needs to change the geometry which implies a more complex 
evaluation of different possible movements of an atom. 
Furthermore, additional processes have to be considered if one aims at the
study of multilayer growth.
Finally, a more realistic description of specific materials requires the use of 
more sophisticated many-body-potentials as, e.g., tight-binding 
RGL (Rosato-Guillope-Legrand) potentials \cite{cleri:prb48}. 


\section{Conclusion}
\label{conclusion}

Our results have confirmed that both microscopic mechanisms, 
strain relaxation {\em and\/} kinetic segregation,
are indeed relevant and crucial for the explanation of
essential features observed in experiments, and that the
experimental findings cannot be explained completely using only 
one of them. 
An equilibrium system with nonzero misfits but otherwise
equivalent particle species displays a checkerboard-like 
mixing of species without the formation of stripes
in the whole range of considered misfits.
On the other hand, a system with zero misfits and different interactions
shows stripe formation but no stable pattern is selected.
The system configurations display segregation into domains
with a characteristic length controlled only by fluctuations
which become very large under close-to-equilibrium conditions.
Moreover, islands growing far-from-equilibrium 
lack the characteristic ramification and asymmetry of material species.

The interplay of, both, different energy barriers and 
misfit induced strain effects together with the effect of kinetics
is needed to explain experimental observations qualitatively.
The presented off-lattice model with nonzero misfits and inter-species interactions
allowed us to reproduce and quantitatively study
the stripe formation as well as the island 
ramification in the segregation regime.

The comparison of results obtained with the off-lattice and the lattice model
show that the presence of chemically induced step edge 
diffusion barriers at A--B interfaces is sufficient for stripe formation.
However, the origin of island edge ramification is more complex.
Ramification was not observed in our lattice model.
A satisfactory treatment of this phenomenon within the framework of a  
lattice gas model will only be possible if the model incorporates 
effectively long-range elastic interactions.

The presented off-lattice model and its modifications allow also for the study of 
related problems appearing in ternary systems.
For instance, it is an interesting open question,
whether the model displays the concentration dependent competition between
alloying and dislocation formation in island growth, which has been 
reported for CoAg/Ru(0001) \cite{thayer:prl86}.
A case of particular interest is that of an anisotropic
substrate which favors the self-assembly of aligned stripes \cite{tober:prl81}. 
Such nanostructures exhibit anomalous magnetic properties \cite{tober:apl77}
which are expected to be relevant in the development of novel storage
devices.


\begin{ack}

This work was supported by Grant Agency of the Czech Republic (Grant No. 202/03/0551).
We further acknowledge support by Deutsche Forschungsgemeinschaft through a research 
grant and Sonderforschungsbereich 410.  

\end{ack}


\end{document}